\documentstyle[epsf]{mn}

\newcommand{\be}{\begin{equation}}
\newcommand{\ee}{\end{equation}}
\newcommand{\bee}{\begin{eqnarray}}
\newcommand{\ene}{\end{eqnarray}}
\newcommand{\et}{\ et\thinspace al.\ }
 
\title{Oscillations of $\alpha\,$UMa and other red giants}

\author[W. A. Dziembowski, D. O. Gough, G. Houdek, R. Sienkiewicz]
       {W. A. Dziembowski$^{1,2,3}$, D. O. Gough$^{1,4}$,
        G. Houdek$^{1}$, R. Sienkiewicz$^{3}$\\
      $^1$Institute of Astronomy, University of Cambridge, Cambridge CB3 0HA,
          UK\\
      $^2$Warsaw University Observatory, Al. Ujazdowskie 4, 00-478 Warsaw,
          Poland\\
      $^3$Copernicus Astronomical Center, Polish Academy of Sciences,
          ul. Bartycka 18, 00-716 Warsaw, Poland\\
      $^4$Department of Applied Mathematics and Theoretical Physics,
          University of Cambridge, Cambridge CB3 9EW, UK}

\date{Accepted 2001 .............
      Received 2000 .............
      in original form ..........}

\begin{document}

\maketitle

\begin{abstract}
There is growing observational evidence that the variability of red giants
could be caused by self-excitation of global modes of oscillation. The most
recent evidence of such oscillations was reported for $\alpha\,$UMa by
Buzasi\et(2000) who analysed space photometric data from the WIRE satellite.

Little is understood about the oscillation properties in red giants. In this
paper we address the question as to whether excited radial and nonradial
modes can explain the observed variability in red giants. In particular, we
present the results of numerical computations of oscillation properties of
a model of $\alpha\,$UMa and of several models of a 2M\,$_\odot$ star in
the red-giant phase.

The red giant stars that we have studied have two cavities that can
support oscillations: an inner core that supports gravity (g) waves and
a surrounding shell that supports acoustic (p) waves. Most of the modes in
the g-mode frequency range are g modes confined in the core; those modes
whose frequencies are close to a corresponding characteristic frequency of a
p mode in the outer cavity are of mixed character and have substantial
amplitudes in the outer cavity. We have shown that such modes of low degree,
$\ell=1$ and $2$, together with the radial (p) modes, can be unstable.
The linear growth rates of these nonradial modes are similar to those
of corresponding radial modes. In the model of $\alpha\,$UMa and in the
2\,M$_\odot$ models in the lower regions of the giant branch, high amplitudes
in the p-mode cavity \hbox{arise only for modes with $\ell=2$.}

We have been unable to explain the observed oscillation properties of
$\alpha\,$UMa, either in terms of mode instability or in terms of
stochastic excitation by turbulent convection. Modes with the lowest
frequencies, which exhibit the largest amplitudes and may correspond
to the first three radial modes, are computed to be unstable if all
effects of convection are neglected in the stability analyses. However,
if the Lagrangian perturbation of the turbulent fluxes (heat and momentum)
are taken into account in the pulsation calculation, only modes with higher
frequencies are found to be unstable.
The observed frequency dependence of amplitudes reported by Buzasi\et(2000)
does not agree with what one expects from stochastic excitation.
This mechanism predicts an amplitude of the fundamental mode about two
orders of magnitude smaller than the amplitudes of modes with orders $n\ge5$,
which is in stark disagreement with the observations.
\end{abstract}

\begin{keywords}
stars: oscillations -- stars: convection -- stars: red giants
\end{keywords}

\section{Introduction}

Buzasi\et(2000) have reported the discovery of oscillations in
photometric data from $\alpha\,$UMa observed with the star camera on the WIRE
satellite. They interpret the observed oscillations as radial modes, and
cautiously suggest that the modes may be excited by the mechanism similar
to that responsible for solar oscillations. The star, however, is
very different from the Sun. Its spectral type is K0$\;$III. Models of
the star's internal structure, and its pulsation properties, suggest that
the star is a red giant. Thus, even if the oscillations are stochastically
excited by turbulence in the outer convective zone, as they are in the Sun,
some important differences between the oscillation properties of
$\alpha\,$UMa and the Sun should be expected.

Variability is a common feature of red giants. There are strong
observationally based arguments that, at least in part, this variability is
due to global pulsations. Edmonds \& Gilliland (1995) proposed radial
or nonradial pulsations as an explanation for the variability they
observed in K giants in the globular cluster 47 Tuc. They found
frequencies between 3 and 6$\,\mu$Hz with amplitudes between
5 and 15\,mmag. Cook\et(1997) analysed photometric data
from red stars in the LMC collected from the microlensing project MACHO, and
reported that the period-luminosity relation comprises several ridges in the
period range of 10\,--\,200-days, which may be interpreted as arising from
radial modes of different $n$. Hatzes and Cochran (1998) reported that there
is strong evidence from radial-velocity data for oscillations in K giants.
Frequencies similar to those found in $\alpha\,$UMa have been
found in a number of other objects. In particular, Hazes and Cochran
quote 11 frequencies of Arcturus (K2$\;$III) in the range
1.4\,--\,6.8$\,\mu$Hz.  Evidence for short-period multimode pulsations in
a number of M~stars has been presented recently by Koen and Laney (2000).

Thus, by being a multimode pulsator $\alpha\,$UMa appears not to be unique
amongst red giants.  But with its high frequencies and low amplitudes it does
represent an extreme case so far, although it is not wholly out of line
with the others. This star is the hottest and the least luminous object amongst
variable red giants. It provides, so far, the best example of possible
red-giant oscillations, but its spectrum is not as
clean as that of the Sun, or of those of other main-sequence stars, or
of white dwarfs. Nevertheless we consider the evidence to be strong enough
to justify new investigations in the theory of red-giant oscillations. So far,
only the modelling of radial pulsations in Miras (see \hbox{Xiong\et1998}
and references therein) and Arcturus \hbox{(Balmforth\et1991)}
has attracted the theorists' attention.
Nonradial oscillations in red giants have been ignored almost entirely.
Here we review the theoretical aspects of this problem, in Section~3, and
provide some numerical examples of the properties of the oscillations of a model
of $\alpha\,$UMa and of models of a 2\,M$_\odot$ star on the red-giant branch.
Some data concerning these models are presented in Section~2.

The most intriguing issue posed by the discovery of oscillations in
red giants is the identification of the mechanism by which they are driven.
Possibilities to consider are
(a) stochastic excitation of linearly stable modes by convection and
(b) self-excitation of linearly unstable modes. We shall speak of
oscillations of case (a) as being solar-like, and case (b), Mira-like.
Our understanding of the excitation mechanism in the Sun
and in Miras is not satisfactory, but the separate association of these
stars with each of the two distinct possible excitation mechanisms
is now generally accepted. In Section~4 we present results of calculations
of radial-mode stability and of the amplitudes in the case of
stochastic excitation.

\section{Selected models}

There are stringent constraints on the parameters for defining models
of $\alpha\,$UMa. The star is bright, and is in a visual binary system.
Accurate spectroscopic data, parallax, and a radius determination
by means of interferometry are available. After considering all the
observational data, Guenther\et(2000) suggested the following values
for the star's global parameters: $M$=4\,--\,5\,M$_\odot$,
$\log(L/L_\odot)=2.5\pm0.05$, $T_{\rm eff}=(4660\pm100)\,$K.
Guenther\et(2000) constructed evolutionary models with masses in this
range and with an initial chemical composition $X=0.727$ and $Z=0.0124$,
which is consistent with the spectroscopic value of [Fe/H] and the
Galactic helium enrichment. They found that only models with
$M\le4.5$M$_\odot$ satisfy the observational constraints.

We have adopted the same initial chemical composition in our model
calculations. Furthermore, we have adopted the same opacity and
equation of state.
For the model of $\alpha\,$UMa we have considered only $M=4\,$M$_\odot$, and
we have adjusted the mixing-length parameter, $\alpha$, to be consistent with
the values of $\log(L/L_\odot)$ and $T_{\rm eff}$ proposed by Guenther\et(2000).

The star $\alpha\,$UMa is a high-mass red giant with a nondegenerate core.
Such stars are very rare. As seen in Fig.~1, the red-giant branch
for $M=4\,$M$_\odot$ is very short; the star spends only 0.5\,My on it,
which is more than two orders of magnitude shorter
than the time spent by a star with a mass of $M=2\,$M$_\odot$, in
which helium ignites in a degenerate core. We have chosen the
model sequence with $M=2\,$M$_\odot$ to illustrate nonradial mode
properties in red giants over a wide range of luminosity. The most
important parameter determining nonradial mode properties is the ratio of 
the mean density of the core to the mean density of the whole star. In 
the sequence we have chosen, this parameter increases by nearly four 
orders of magnitudes between the bottom and the top of the giant branch.
We have considered four models for the 2\,M$_\odot$ sequence, calculated with
the same values of $X$, $Z$ and $\alpha$ as those for the model of
$\alpha\,$UMa. The locations of the selected models on the evolutionary tracks
are indicated in Fig.~1; the parameters characterizing these models are listed
in Table~1.

\begin{table*}
\caption{Some parameters of the models used in this
work; $M_{\rm c}$ is the core mass including the hydrogen burning shell;
the subscript bc denotes bottom of the convective envelope. }
\begin{tabular}{ccccccccccc}
\hline Model &  $M/M_\odot$ & age(Gy) & $\log T_{\rm eff}$ & $\log
L/L_\odot$ & $R/R_\odot$ & $\log T_{\rm c}$ & $\log\rho_{\rm c}$ & $M_{\rm c}$ &
$r_{\rm bc}/R$ & $M_{\rm bc}/M$\\ \hline M$_{\alpha}$ & 4 & 0.143 & 3.6993
& 2.50 & 27.21 & 7.924 & 4.187 & 0.1083 &
  0.3259 &  0.4300\\
 M21 & 2 & 0.904 & 3.6915 & 1.50 & 7.77 & 7.753 & 4.766 & 0.1145 & 0.1481 &
 0.1869\\
 M22 & 2 & 0.934 & 3.6596 & 2.00 & 15.99 & 7.776 & 5.220 & 0.1397 & 0.0509 &
 0.1527\\
 M23 & 2 & 0.963 & 3.6240 & 2.50 & 33.50 & 7.780 & 5.555 & 0.1705 & 0.0264 &
 0.1755\\
 M24 & 2 & 0.972 & 3.5848 & 3.00 & 71.32 & 7.835 & 5.752 & 0.2017 & 0.0124 &
 0.2042\\
\hline
\end{tabular}
\end{table*}

\begin{figure}
\epsfxsize=8cm \epsffile{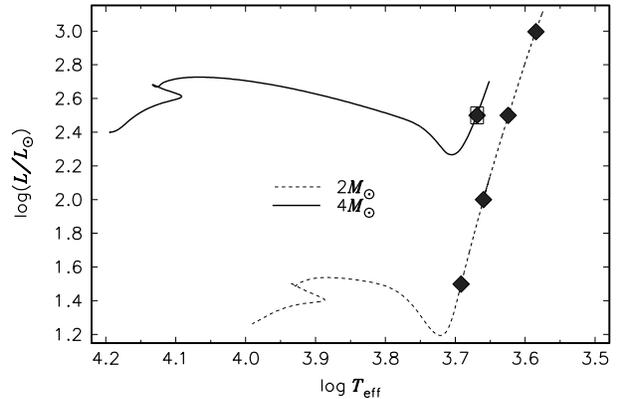} \caption{ H-R diagram
showing evolutionary tracks of models with masses of 2 and 4
M$_\odot$. Models selected for pulsation analyses are indicated by
diamonds. The box around the symbol for the 4\,M$_\odot$ star indicates
the uncertainty in locating $\alpha\,$UMa in the diagram (see Table 1).}
\end{figure}

\section{Nonradial modes of red giants}

Guenther\et(2000) considered only radial modes as potential
candidates for explaining the peaks in the $\alpha\,$UMa
frequency spectrum determined by Buzasi\et(2000). They noticed that
the frequencies of these peaks are much lower than the buoyancy frequency
deep in the star, and presumed that a nonradial interpretation would appear
to imply that the modes are g modes of high radial order.
According to their estimate, the separation between the cyclic frequencies of
consecutive g modes of like degree is of the order of 0.1$\,\mu$Hz, and
consequently, they argued, the spectrum could not be resolved into individual
modes. They did not, however, explain why radial modes should stand above this
quasi-continuum, which would appear to be necessary for explaining Buzasi's
observations.
Moreover, they failed to point out that a g mode that resonates at a
corresponding (i.e., same value of $\ell$) characteristic p-mode frequency
of the outer acoustic cavity can have a particularly large amplitude at the
surface.

\subsection{General properties}

The basic properties of nonradial oscillations in highly evolved stars were
determined in the 1970s (Dziembowski, 1971, 1977; Osaki, 1977).
However, the objects of interest in those early works were stars in the
Cepheid instability strip. To the best of our knowledge there is
only one paper devoted to the theory of nonradial oscillations in red
giants. It is a short note by Keeley (1980), in which a crude
estimate of mode trapping was made. The nonadiabatic effects, which
are very important in this context, were ignored.

The differences in the nonradial mode properties between red and yellow
giants are a consequence of the different depths of the convection zones.
The formalism for calculating linear modes in these two types of
star is the same. Here we provide only an outline of
the formalism developed by Dziembowski (1977), which was recently
recalled in some detail by Van Hoolst\et(1998, hereafter VDK).
We intend to apply this formalism to low-degree modes ($\ell=1,2$),
in a cyclic-frequency range starting somewhat below the fundamental
radial-mode frequency and extending up to the acoustic cut-off
frequency, $\nu_{\rm ac}$, in the photosphere ($r=R$). In our model of
$\alpha\,$UMa, $\nu_{\rm ac}\simeq27\mu$Hz.
Buzasi\et(2000) reported peaks in the power spectrum located above
our value of $\nu_{\rm ac}$, which evidently cannot easily be interpreted
in terms of strongly trapped acoustic modes.

The starting point of our discussion is an asymptotic solution, for large
order $n$, of the nonadiabatic wave equation, which is valid in the radiative
interior. In this approximation, any perturbed scalar parameter may be
expressed with respect to spherical polar coordinated ($r, \theta, \phi$) in
the following form:
\be
q(r,\theta,\phi,t)=A(r)\left[{\rm e}^{{\rm i}\Phi(r)} +{\rm e}^{-{\rm i}
\Phi(r)}\right]Y_\ell^m(\theta,\phi){\rm e}^{i\omega_{\rm c} t},
\ee
in which $t$ is time. The amplitude, $A$, is a slowly varying function
of $r$. The rapid variations are described by the phase $\Phi$, which for
stars with radiative cores may be written in the form
\be
\Phi(r)=\int_0^rk\,{\rm d}r-{1\over2}(\ell+1)\pi.
\ee
The general expression for the radial component, $k$, of the wave number of
high-order modes in the gravity-wave cavity may be found in VDK.
The quantity $\omega_{\rm c}=\omega -{\rm i}\gamma$ (with $\omega>0$) is the
complex eigenfrequency.
We focus our attention on predominantly oscillatory modes, i.e. modes with
a growth rate $\gamma$ satisfying the condition $\vert\gamma\vert\ll\omega$.
We also specify $q$ to be the relative Lagrangian perturbation to the
pressure, $\delta p/p$.

If the radiative energy losses are regarded as being small, there is a
simple expression for the radial wave number far from the edges of the cavity:
\be
k\simeq{\sqrt{\ell(\ell+1)}\over\omega}{N\over r}
       [1+{\rm i}({\cal D}+{\gamma\over\omega})],
\ee
where
\be
N=g\sqrt{\left({{\rm d}\rho\over{\rm d}p}-{\rho\over\Gamma_1p}\right)}
\ee
is the buoyancy (Brunt-V\"ais\"al\"a) frequency, $g$ is the local gravitational
acceleration, $\rho$ is density,
$\Gamma_1=(\partial\ln\,p/\partial\ln\,\rho)_{\rm ad}$ is the first adiabatic
exponent, and
\be
{\cal D}={\ell(\ell+1)\over 8\pi\omega^3} {gL_{\rm r}\over r^4p}
         {\nabla_{\rm ad}\over\nabla}(\nabla_{\rm ad}-\nabla)\,,
\ee
in which $L_{\rm r}(r)$ is the total rate at which radiant energy crosses
a sphere of radius $r$, $\nabla={\rm d}\ln\,T/{\rm d}\ln\,p$ and
$\nabla_{\rm ad}=(\partial\ln\,T/\partial\ln\,p)_{\rm ad}$.
The quantity ${\cal D}$ is a measure of the radiative energy loss,
which is the only nonadiabatic effect we consider in this cavity. Here,
$\nabla_{\rm ad}>\nabla$ is always satisfied; hence ${\cal D}>0$. To the
same approximation, the amplitude is given by
\be
A(r)\propto{V\over r^3\sqrt{k\rho}}\,,
\ee
with $V=gr\rho/p$.
The approximate expressions for the wave number (equation~3) and for the
amplitude (equation~6) are valid only if ${\cal D}\ll 1$. For models
M23 and M24 (see Table~1) the computations suggest that ${\cal D}\gg 1$
in certain layers inside the star, at least for some modes considered in the
calculations. However, for these models we use this approximation only in the
outer layers of the asymptotic region, where the approximation is satisfied.
The maximum value of
${\cal D}$ in our model of $\alpha\,$UMa for the lowest-frequency quadrupole
($\ell=2$) mode is 0.2. As in all red-giant models, that maximum occurs
within the shell source. Even if ${\cal D}$ is small, consequences of
radiative losses may still be very important for the wave properties, because
if $N/\omega$ is large we may still have $\Phi_{\rm i}\equiv\Im(\Phi)\gg1$.
Equation~(1) describes a superposition of an outward (first term on the rhs)
and inward propagating gravity wave, where the direction of propagation is
the direction of the group velocity.

Let us concentrate now on the oscillations in the outer, acoustic cavity.
Moreover, let us assume that the approximation for $q$ given by equation~(1)
is valid in an interval $[0,r_{\rm f}]$ of the g-mode cavity. In this interval
we can neglect the derivative of $A$ with respect to $r$ in the calculation
of the derivative of $q$. Thus, at $r=r_{\rm f}$ we have approximately
\be
{1\over q}{\partial q\over\partial r}={\rm i}k{\exp({\rm i}\Phi)-
\exp(-{\rm i}\Phi)\over\exp({\rm i}\Phi)+\exp(-{\rm i}\Phi)}\,,
\ee
which provides a boundary condition for the
numerical solution of the equation for the linear nonadiabatic oscillations
in the interval $[r_{\rm f},R]$, which contains the very outer layers of the
g-mode cavity in which the asymptotics breaks down, the entire surrounding
acoustic cavity and the evanescent zone between.
The lhs of equation (7) depends on $\ell$ and on the value of
$\omega_{\rm c}$. The dependence on $\omega_{\rm c}$ is relatively weak
in comparison with the explicit dependence of the rhs of equation~(7).

Assuming $\Phi_{\rm i}\gg1$, expression (7) simplifies to
\be
{\partial q\over\partial r}\simeq -{\rm i}k q\,,
\ee
which is valid for the case when the wave is effectively dissipated
on its way towards the centre. The energy loss may be
overcompensated by the driving operating in the outer layers if
the wave amplitude $A(r_{\rm f})$ is small, i.e., if the mode trapping
in the acoustic propagation zone is severe. Indeed, nonradial
modes with growth rates $\gamma$ similar to those of radial modes were
found in models of Cepheids (Osaki, 1977) and of RR Lyrae
(Dziembowski, 1977). In these two independent papers, equation (8) was used
for the inner boundary condition. Such modes were named in VDK as
S(trongly) T(rapped) U(nstable).
We must emphasize, however, that even when the amplitude in the outer
acoustic zone is relatively large, according to VDK the oscillations typically
have 80\,\% of their energy in the inner, g-mode cavity.
In the model of the RR Lyrae star considered by VDK, STU modes
were found only with $\ell>4$. We shall see that in red giants
STU modes may exist also with $\ell$ as low as unity. The frequency
separation between consecutive low-degree STU modes is similar to
the separation between consecutive radial modes.

The STU modes are true eigensolutions of the nonadiabatic
oscillation equations for the whole star. Indeed, the boundary
conditions (7) and (8) are equivalent for unstable modes
because  if $\gamma>0$ we have $\Im(k)>0$ throughout the interval
$[0,r_{\rm f}]$, and $\Phi_{\rm i}$ does not change sign.
If ${\cal D}\ll1$, this conclusion follows immediately from equation~(3)
although, in fact, it is true also for any  ${\cal D}\ge0$
(see e.g. Dziembowski, 1977).  For stable modes, the situation is more involved.
If a solution with $\gamma<0$ is found subject to the inner boundary
condition (8), then the solution must always be checked to determine whether
it satisfies the inequality $\Phi_{\rm i}\gg1$. Actually, this inequality is
rarely satisfied. Let us note that with the usage of equation (7) we
assume maximum energy losses. When we use equation (7) with a properly
calculated phase, instead of equation (8), we may find unstable modes.
However, for such modes the growth rates are typically much smaller than
those of radial modes with similar frequencies.

A dense spectrum of weakly unstable modes with $\ell=1$ and 2 was
found for the RR Lyrae model considered by VDK.
We shall discuss in the next Section the problem of mode stability,
and we shall see that it is actually far from being solved. Fortunately,
whatever the mechanism responsible for the excitation of the modes, it
should operate in the layers where the radial eigenfunctions do not
depend on $\ell$. We shall take advantage of this property in our
discussion of the relative chances of nonradial or radial
modes being excited.

It seems to be not unreasonable to assume that if in a certain
frequency range unstable modes of various degrees exist, their
chances of being excited are related to the growth rates $\gamma$.
The growth rate may be expressed in terms of the work integral, $W$,
and the mode inertia, $I$, through the well-known relation
(see e.g. \hbox{Unno\et1989)}
\be
\gamma ={W\over 2\omega I}.
\ee
The generic expression for the work integral is
\be
W=\int{\rm d}^3{\bmath x}\,\rho[-T\nabla_{\rm ad}\Im(q^*\delta s)]
 +\int{\rm d}^3{\bmath x}\,\Im\left ({\delta\rho\over\rho}^*
                        {\delta p_{\rm t}\over p_{\rm t}}\right )\,,
\ee
where $s$ is the entropy per unit mass and $p_{\rm t}$ (the so-called
turbulent pressure) is the $rr$-component of the Reynolds stress tensor
$T_{ij}\equiv\overline{\rho u_iu_j}$ ($\bmath u$ is the turbulent velocity
field and the overbar denotes an ensemble average). The asterisk denotes
complex conjugate. In this expression we neglect the contribution from the
anisotropy of the Reynolds stress tensor, which is small compared with the
isotropic component. But in this section we neglect convection dynamics in
the model computations: in the evaluation of the work integral the second
term of the rhs of equation~(10) is neglected.
The mode inertia is defined as
\be
I=\int {\rm d}^3{\bmath x}\,\rho\vert{\bmath\xi}\vert^2,
\ee
with ${\bmath\xi}$ representing the displacement eigenfunction.
The integrals are over the entire volume of the star.
The inertia $I$ enters also into the expression for the amplitudes of
stochastically excited modes (see equation~21). Another quantity in the
expression for the amplitudes is the energy supply rate $P_Q$ injected into
the modes by the turbulent convection, and which we assume to be generated
predominantly by the fluctuating Reynolds stresses (see next Section).

There are important differences between radial and nonradial oscillation
properties below the acoustic propagation zone of the nonradial modes.
These differences are reflected in the values of $I$. If $I$ is large, the
largest contribution to the work integral may arise
in the gravity-mode propagation zone, where the asymptotic approximation is
applicable. In the gravity-wave propagation zone we have adopted for the
Lagrangian specific entropy perturbation the expression
\be
\delta s=2\,{\rm i}\,c_p\left({\partial\ln T\over\partial\ln\rho}\right)_p
 \left({{\rm d}\ln p\over{\rm d}\ln\rho}-{1\over\Gamma_1}\right){\cal D}\,q\,,
\ee
in which $c_p$ is the specific heat at constant pressure. It follows, for
example, from equation~(19) of VDK in the weakly nonadiabatic limit,
that is, when $\vert{\cal D}\vert\ll1$.
Substituting equations (1) and (12) into equation (10) we obtain
\be
W_{\rm g}=-2C_{\rm f}\omega^2\int_0^{\Phi_{\rm r,f}}h{\cal D}d\Phi_{\rm r},
\ee
where $\Phi_{\rm r}\equiv\Re(\Phi)$ and $C_{\rm f}$ is a real positive constant
which is obtained from the eigenfunctions calculated numerically for
$r>r_{\rm f}$, and
\be
h=\exp(2\Phi_{\rm i})+\exp(-2\Phi_{\rm i})\,.
\ee
The oscillatory term, proportional to $\cos(2\Phi_{\rm r})$, in the integrand
of $W_{\rm g}$ has been ignored, which is consistent with the asymptotic
approximation.

An expression for $\bmath\xi$ in terms of $q$ can be calculated in the
adiabatic approximation. The result is
\be
{\bmath\xi}=-{r\over V}\left[q\,{\bf e_{\rm r}}+{r^2\over\ell(\ell+1)}
             \nabla_{\rm h}
\left({\partial q\over\partial r}\right)\right]\,,
\ee
where ${\bf e_{\rm r}}$ is a unit vector in the radial direction and
$\nabla_{\rm h}$ is the horizontal component of the gradient operator.
From equation~(3) we conclude that, to a first approximation, the contribution
of the radial displacement to $I$ can be neglected if $N\gg\omega$.
Thus, from the asymptotic interior we obtain the following contribution
to the modal inertia $I$ :
\be
I_{\rm g}=C_{\rm f}\int_0^{\Phi_{\rm r,f}}h\,{\rm d}\Phi_{\rm r}.
\ee

We normalize the relative rms radial component of the mode displacement
to unity at the stellar surface. The coefficient $C_{\rm f}$, which is
a function of $\omega$, exhibits minima separated by nearly the same
interval in frequency as the minima in the radial modes. This is a
manifestation of the trapping properties of the acoustic cavity, which
are purely dynamical and result from a resonance between the two cavities.
The spectrum of g modes in the inner cavity is so dense that there is always
a g-mode-like oscillation whose frequency resonates with a p mode in the
outer cavity, such that the amplitudes in both cavities are similar.
All other g modes are confined to the central g-mode cavity, and have
very low amplitudes in the outer layers of the star.
Mode trapping is influenced also by the behaviour of the factor $h$
(see equation 14), which is determined by nonadiabatic effects.
For STU modes we have $h\approx\exp(2\Phi_{\rm i,1})\gg1$, which is
a sharply increasing function of $r$. Thus, $I_{\rm g}$ is negligible and
$W_{\rm g}$ may be evaluated as the rate of wave losses:
\be
W_{\rm g,w}=-\left(r^2p\int\Im(\xi_r^*q)\sin\theta\,{\rm d}\theta
{\rm d}\phi\right)_{\rm f}=-C_{\rm f}\omega^2h_{\rm f}\,.
\ee
For all other modes we have to
use equations (13) and (16) to evaluate the contributions $W_{\rm g}$ and
$I_{\rm g}$. Equation (3) implies that for stable modes $h(r)$ has
a maximum in the layer in the star in which ${\cal D}=-\gamma/\omega$.
Thus, $I_{\rm g}$ may be a significant, and is often the dominant contribution
to $I$. Let us note that the values of $I_{\rm g}$ and $W_{\rm g}$ depend on
$\gamma$, and consequently on the nonadiabatic processes operating in these
layers. This means that uncertainties in the computation of the nonadiabatic
effects are to some degree reflected in the values of $I_{\rm g}$ and
$W_{\rm g}$.  Damping in the outer layers reduces the effect of mode trapping.

\subsection{Application to $\alpha\,$UMa}

In our code for computing nonradial nonadiabatic oscillations
(Dziembowski, 1977) we set the Lagrangian perturbation of the turbulent
fluxes (heat and momentum) to zero, and we ignore the turbulent pressure in
the equilibrium model. With this treatment, all radial modes are found to be
unstable.

\begin{figure}
\epsfxsize=8cm
\epsffile{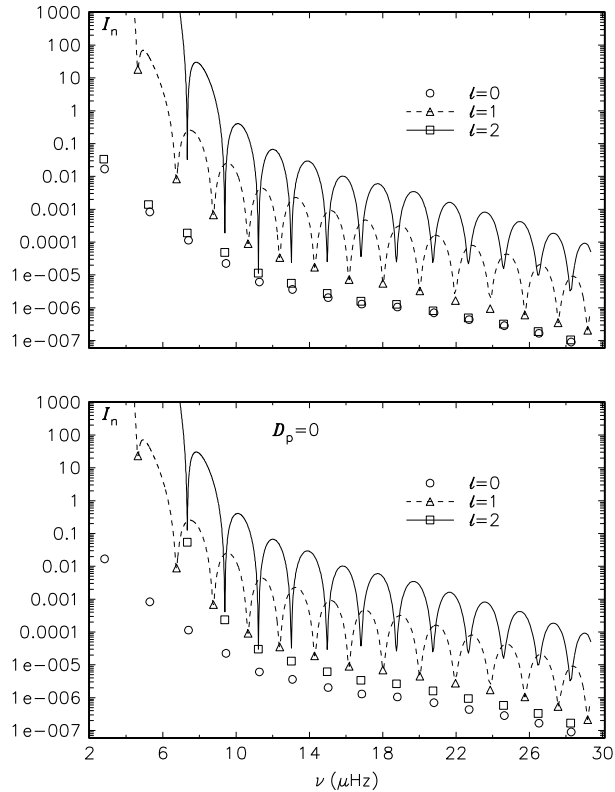}
\caption{ Modal inertia in units of $3MR^2$, plotted against frequency.
The eigenfunctions are normalized such that at the surface
${\xi_{\rm r}}=R\,Y_l^m(\theta,\phi)\exp({\rm i}\omega_{\rm c} t)$.
Individual nonradial modes are not resolved. The symbols are
displayed only for those modes that are locally most trapped.
$D_{\rm p}=0$ means that all nonadiabatic effects in the outer layers
are ignored.}
\end{figure}

\begin{figure}
\epsfxsize=8cm \epsffile{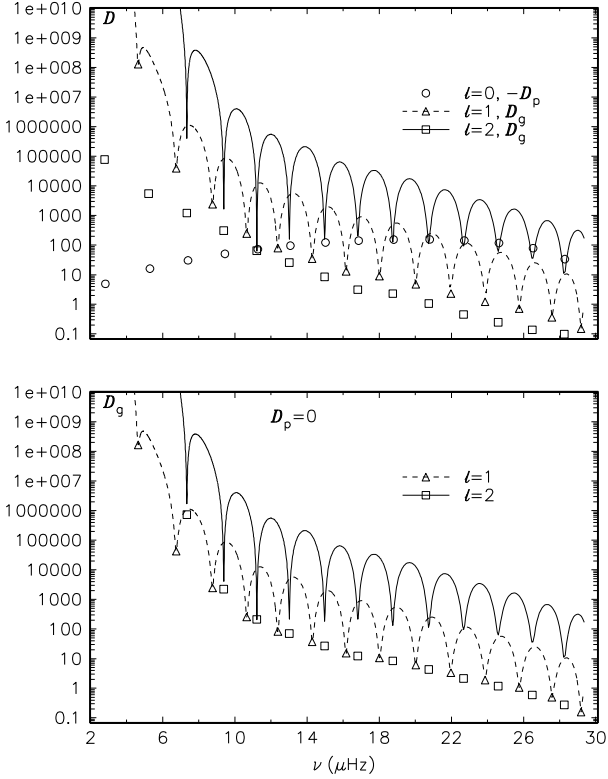} \caption{ Contributions to the
energy dissipation rate from the g-mode propagation zone in units
of the stellar luminosity, $L$ . In the upper panel the energy gain
rate for radial modes is plotted with open circles. See caption
of Fig.~2 for further information.}
\end{figure}

In Fig.~2 we show the behaviour of the normalized mode inertia
$I_{\rm n}=I/3MR^2$ as a function of the cyclic frequency $\nu=\omega/2\pi$
for our model of $\alpha\,$UMa. The choice of normalization is not
important here, except that all modes are assumed to have the same surface
amplitude of radial displacement. There are two sequences of model results:
in the first sequence (upper plots) we calculated $h$ with $\gamma$ obtained
from our code; in the second sequence we suppressed all
nonadiabatic effects where $r>r_{\rm f}$. A comparison allows us to assess
some of the consequences of the uncertainties of the physics in the
convective zone.

Symbols are used to denote the nonradial modes that are most trapped in the
acoustic cavity. For the $\ell=2$ sequence the minima in $I_{\rm n}$ almost
coincide with the radial-mode frequencies, while the minima for $\ell=1$ are
located roughly half-way between
the $\ell=0$ and $\ell=2$ minima. The positions of these minima resemble the
positions of modes in the whole-disc spectra of solar oscillations. There
are many more modes with $\ell=1$ and $\ell=2$ than are depicted by the symbols.
The frequency separation $\Delta\nu$ between nonradial modes of consecutive
order $n$ is indeed very small. It may be evaluated from the asymptotic
formula
\be
{\Delta\nu\over\nu}=2.4\times10^{-4}{\nu\over\sqrt{\ell(\ell+1)}}\,,
\ee
where $\nu$ is expressed in $\mu$Hz. The numerical constant is specific
to the model. At $\ell=1$ and $\nu=2.8\mu{\rm Hz}$ the value of
$\Delta\nu$ is about 0.0013$\,\mu$Hz, much less than that found by
Guenther\et(2000).

There is no substantial difference between the trapping pattern of the two
sequences, except for the differences in the depths of the minima, particularly
those of the $\ell=2$ modes. Greater driving in the outer layers results in
deeper minima. If there is net damping in the outer layers, %($D_{\rm p}<0$),
as in the case of solar oscillations, the minima are shallower than in the
adiabatic approximation.

In Fig.~3, we plot the rate of energy dissipation,
$D_{\rm g}\equiv-\omega W_{\rm g}$, in the asymptotic interior for the same
two sequences of modes. In addition, in the upper panel, we show the total
energy gain rate, $-D_{\rm p}\equiv-\omega W$, for radial modes. The total
driving rate for the nonradial modes is given approximately by
$\gamma\simeq(D_{\rm p}(\nu)+D_{\rm g})/2\omega^2I$,
because significant nonadiabatic effects may arise only either near the
surface\,--\,and then they are $\ell$-independent\,--\,or in the deep interior
where the g-mode asymptotics applies
Some nonradial modes with
frequencies larger then 12$\,\mu$Hz are found to be unstable:
if the radial modes with $\nu>12\,\mu$Hz are indeed unstable, then there
are also some unstable low-degree nonradial modes.

When $\ell=2$ the modes that are most trapped are detached from the
remaining modes, except for the one at $\nu\simeq10\,\mu$Hz.
Except for this particular mode, all the other modes satisfy
$\Phi_{\rm i,f}>1$. Thus, the unstable $\ell=2$ modes are STU modes,
and their growth rates are nearly the same as those of the corresponding
radial modes.  All the unstable $\ell=1$ modes have $\Phi_{\rm i,f}\ll1$,
and the trapping effect is less severe. The inertiae of the most trapped
$\ell=1$ modes are always significantly larger than those of the closest
radial mode (see Fig.~2).

\subsection{``Unstable'' low-degree modes in {\bf 2\,M$_\odot$} red giants}

In Table~2 we compare some characteristics of modes of the $\alpha\,$UMa
model, M$_\alpha$, and the modes of models of 2\,M$_\odot$ red giants that
are found to be unstable with our code. For the models M21 and M22 we find
more-or-less similar properties to those of the M$_\alpha$ model. Strong
trapping occurs only for $\ell\ge2$, and the nonradial modes are unstable for
the higher $n$. Note that $n$ is the radial order only for $\ell=0$
(for $\ell>0$ it is the number of nodes in the acoustic cavity of the radial
component of the displacement eigenfunction).
For all cases the upper limit of the unstable
range is determined by the acoustic cut-off frequency.

\begin{table*}
\caption{``Unstable'' low-degree modes in red-giant models; $\nu_{\rm ac}$ is
the cyclic acoustic cut-off frequency computed for an Eddington grey
atmosphere; $\Pi_{0,1}$ is the period of the
fundamental radial mode; for $\ell>0$, $n$ indicates the range of consecutive
modes that are most trapped.}
\begin{tabular}{ccccccccc}
\hline Model&$\nu_{\rm ac}$ &$\Pi_{0,1}$& &$n$-range& & & $\nu$-range ($\mu$Hz)
 &\\
 &$\mu$Hz& days&$\ell=0$&$\ell=1$&$\ell=2$&$\ell=0$&$\ell=1$&$\ell=2$\\
\hline
M$_{\alpha}$ & 26.6 & 4.10&1-14&8-14&4-12 &2.82-26.3&16.2-27.7&15.0-28.2\\
M21 & 178. & .919 & 1-18 & 6-18 & 6-18 &12.6-166.&59.5-162.&62.5-166.\\
M22 & 43.4 & 2.65 & 1-13 &  5-14 & 3-13 &4.36-40.3&16.3-41.9&11.5-40.3\\
M23 & 10.2 & 7.74 & 1-9  &  2-10 & 1-12 &1.50-9.30&2.37-9.81&1.49-10.1\\
M24 & 2.25 & 24.1 & 1-6 &   1-6 &  1-6  &.480-2.03&.397-1.92&.483-2.01\\
\hline
\end{tabular}
\end{table*}

In the more luminous giants (models M23 and M24) STU modes are found even
for $\ell=1$. In Fig. 4, we show how instability of the most strongly trapped
modes increases with stellar luminosity.

\begin{figure}
\epsfxsize=8cm \epsffile{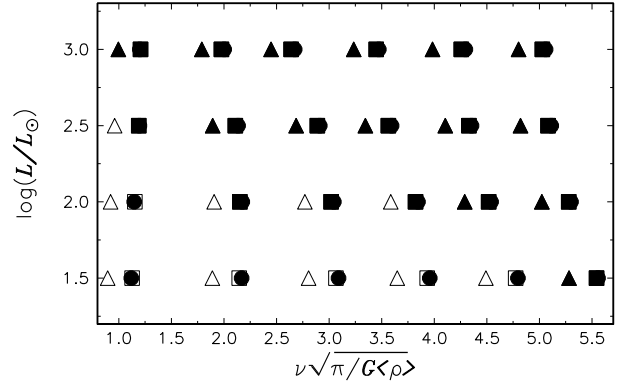} \caption{Occurrence of
instability in the sequence of 2\,M$_\odot$ giants of
nonradial mode of low degree $l$ and low radial order $n$
that are most strongly trapped in the outer acoustic cavity.
The abscissa is the dimensionless frequency $\sqrt{\pi/G<\rho>}\nu$.
Full symbols denote unstable modes, open symbols stable modes.
Circles, triangles, and squares denote $\ell=0, 1$, 2 modes, respectively}
\end{figure}

The relative frequencies of the most strongly trapped modes of the models
considered here are different from those of the RR Lyrae star model considered
by VDK and of RR~Lyrae stars in general. In red giants the most strongly
trapped $\ell=1$ modes are located between the $\ell=0$ and $\ell=2$ pairs,
whose frequencies are nearly coincident.
There is a similarity with solar p modes, although in the case we have studied
here the frequencies of the strongly trapped $\ell=1$ modes are somewhat
closer to the higher-frequency even-degree pair. In RR Lyrae stars, on the
other hand, the frequencies of the most strongly trapped $\ell=1$ modes are
actually closer to the radial eigenfrequencies than are the frequencies of
the most strongly trapped $\ell=2$ modes.

These differences between RR Lyrae stars and red giants are
related to red giants having much deeper convective zones. 
The differences
are reflected in the different behaviours of the
Brunt-V\"ais\"al\"a frequency, $N$. In Fig.~5 we compare $N$ and
the Lamb frequencies $L_1$ and $L_2$ in the envelope of an RR
Lyrae model with those in the envelope of model M21. A deeper
convective zone is associated with a wider evanescent zone separating the 
p-mode and g-mode propagation zones, and hence there is a possibility of 
more efficient trapping. This is why we find STU modes of low
degree in red-giant and not in RR Lyrae models. Exceptionally poor
trapping of the $\ell=2$ modes in the frequency
range of the first two radial modes is due to the narrowness of the
evanescent zone.

\begin{figure}
\epsfxsize=8cm \epsffile{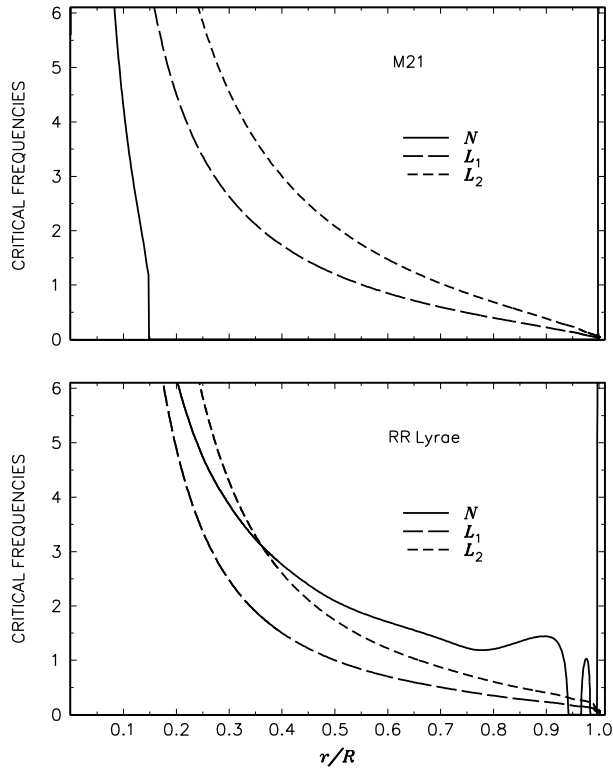} \caption{Lamb frequencies
$L_\ell=\sqrt{\ell(\ell+1)}c/r$ (for $\ell=1$ and 2) and the 
Brunt-V\"ais\"al\"a frequency $N$ in the M21 model and in a representative 
RR Lyrae star model. The latter is characterized by the following
parameters: $M=0.67$M$_\odot$, $Y_0=0.243$, $Z=0.001$, $Y_{\rm c}=0.17$
(helium abundance in the core), $\log(L/L_\odot)=1.717$, $\log
T_{\rm eff}=3.822$. The ordinate scale is dimensionless, and
corresponds to angular frequencies measured in units of
$\sqrt{4\pi G\!<\rho\!>}$, which corresponds to the dimensionless
cyclic frequencies of Fig.~4. In these units the frequencies of
the first two radial modes in the RR Lyrae model are 1.82 and
2.46.}
\end{figure}

\section{Excitation mechanisms}

There is little doubt that the interaction between pulsation and
convection plays an essential role in red-giant pulsation, and that
the approach adopted by us to obtain the results reported in the previous
sections is inadequate. The driving agent that caused instability of the
radial modes is the same as that suggested first by Ando and Osaki (1975)
in an attempt to explain solar p-mode excitation in the Sun, and is
artificial. It is easy to understand why:
at the photosphere, where the energy is carried mostly by radiation, the
flux perturbation is negative in the high-temperature phase of the
pulsation cycle. This is a result of the steep increase of the opacity
with temperature in the outer layers. The fraction of the energy carried
by convection increases rapidly inwards. Since, {\it by assumption}, the
convective flux remains unperturbed, the energy is forced to be captured by
the photospheric layers, and the putative heat engine works. This
phenomenon is sometimes called convective blocking, which is
confusing because what actually blocks the heat flux is the opacity
variation. However, there is no physical justification for the
neglect of the perturbed convective heat flux and Reynolds stresses.
Indeed, pulsational modulation of the convectively unstable stratification of
the star is bound to modulate the convective dynamics, and dominate the driving
or damping in regions where the convective fluxes dominate in the equilibrium
state.

Effects of convection on the stability of radial pulsations in cool stars
have been investigated since the early 1970s (see e.g.
\hbox{Xiong\et1998}; \hbox{Houdek, 2000)}.
Recent efforts have focused mainly on Mira stars and the Sun.
According to the calculations of Xiong\et(1998), low-order radial modes
of Mira models are unstable, whereas those of orders
$n>4$ were always found to be damped (see also the work by
Balmforth\et(1991) on Arcturus). In a study of p-mode stability in the
Sun by Balmforth (1992a), all modes have been
found to be stable. Balmforth used in his calculations
Gough's (1976, 1977) nonlocal, time-dependent mixing-length model for
convection, improving on the code used by Baker \& Gough (1979) to study
RR Lyrae stars by incorporating the Eddington approximation to
radiative transfer for both the equilibrium structure and the pulsations.
Houdek\et(1999) applied these calculations to solar-type stars, and estimated
amplitudes of intrinsically stable stochastically excited radial
oscillations in stars with masses between 0.9\,M$_\odot$ and 2.0\,M$_\odot$
close to the main sequence.

\subsection{Linear stability of radial modes in $\alpha\,$UMa}

Here we apply Balmforth's (1992a) treatment of pulsation to a model
of $\alpha\,$UMa. In particular, we include turbulent pressure in the
equilibrium model, and the stability analysis includes the Lagrangian
perturbations of the convective heat and momentum fluxes.
We use an envelope model calculated with the surface parameters of 
model M$_\alpha$ given in Table~1, and an atmosphere using the $T$-$\tau$ 
relation of model C of Vernazza, Avrett \& Loeser (1981).
The value of the
mixing-length parameter was adjusted such as to reproduce the same
depth of the convective zone as was obtained from the evolutionary
computation. The nonlocal treatment of convection introduces two more
parameters, $a$ and $b$, which characterize respectively the spatial
coherence of the ensemble of eddies contributing to the total heat and
momentum fluxes and the extent over which the turbulent eddies experience
an average of the local stratification. Theory suggests approximate values
for these parameters, but it is arguably better to treat them as free.
Roughly speaking, the parameters control the degree of `nonlocality' of
convection; low values imply highly nonlocal solutions, and in the limit
$a,b\rightarrow\infty$ the system of equations reduces to the
local formulation (except near the boundaries of the convection zone, where
the local equations are singular).

The energy dissipation rate $D_{\rm p}$ of radial p modes was calculated as a
continuous function of oscillation frequency by relaxing the inner dynamical
boundary condition. The results shown in the upper panel of Fig.~6 were
obtained for two sets of the nonlocal convection parameters $a$ and $b$.
The choice of these parameters is important at high frequencies where
unstable frequency ranges are found. At low frequency, covering radial orders
up to $n=5$, all modes are found to be stable for both sets of the
$a$ and $b$ parameters. The values of $\vert D_{\rm p}\vert$ are significantly
higher than the values of $-D_{\rm p}$
shown in the upper panel of Fig. 3.
This clearly indicates that by neglecting the perturbed convective fluxes we
ignore the dominant contribution to the damping.

\begin{figure}
\epsfxsize=8cm
\epsffile{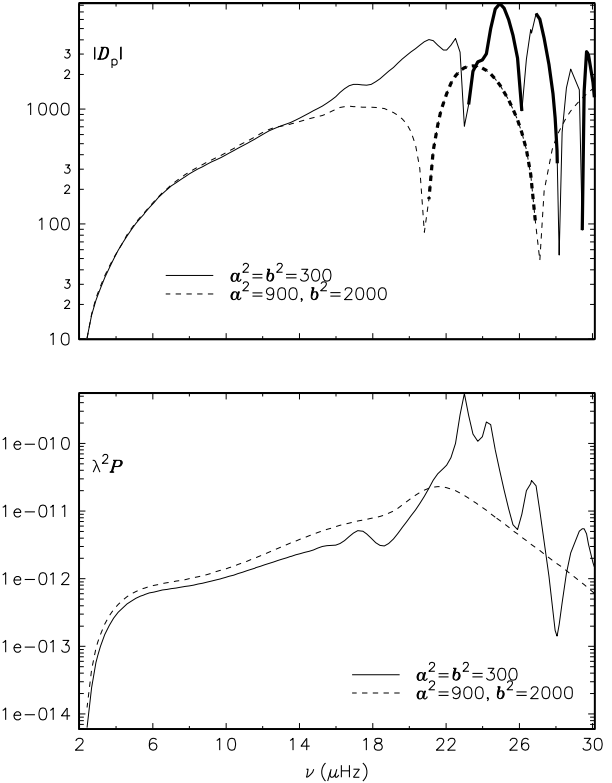}
\caption{ Absolute values of energy dissipation rates,
$\vert D_p\vert$ (top) and energy generation rate $P\lambda^2$ (bottom)
($P$ is the rate of energy injected into the modes by the fluctuating
Reynolds stresses and $\lambda$ is defined by equation~22). The energy
generation rates are expressed in units of the solar luminosity.
Two sets for the convection parameters $a$ and $b$ were used. Thick
curves in the upper panel indicated the frequency range where radial modes
are found to be unstable ($D_p<0$).}
\end{figure}

The results shown in Fig.~6 are applicable also to nonradial modes,
because virtually all the contribution to $D_{\rm p}$ arises in the upper
layers where the value of $\ell$ has little influence.
However, damping effects in these layers have consequences in the
deep interior. They change $\gamma$, and hence the amplitude behaviour in
the g-mode propagation zone (see equation~3).
We have seen in Section 3.2 (Figs. 2 and 3) that ignoring driving
effects in these layers reduces the trapping. Adding damping there would
reduced it further. Larger inertiae imply lower amplitudes for
stochastically exited modes, and indeed we should not expect
a detection of stochastically excited nonradial modes in giants.

\subsection{Amplitudes of stochastically driven radial modes}

The amplitudes of intrinsically stable stochastically driven radial modes
were estimated in the manner of Houdek\et(1999):
\be
V_s=\sqrt{P_Q\over2\eta I_\omega},
\ee
where here $P_Q$ is the noise generation rate injected into a mode through
the fluctuating Reynolds stresses, the expression for which we adopted
from Balmforth (1992b) (see also \hbox{Houdek\et1999)}. The damping rate is
$\eta=D_{\rm p}/2 I\omega^2=-\gamma$, and $I_\omega=IR^{-2}$ in our notation.
For radial modes the total energy dissipation rate $D$ is $D_{\rm p}$.
The linear stability analysis also
provides the parameter $\lambda$, which is the ratio of the relative
luminosity to the relative velocity amplitude, computed at the
surface (i.e. outermost meshpoint) of the star. The bolometric relative
luminosity amplitude then becomes
\be
{\delta L\over L}=\lambda{\delta R\over R}=\lambda{V_s\over\omega R},
\ee
and from equation~(21) we obtain,
\be
{\delta L\over L}=\lambda\sqrt{P\over I_{\rm n} D}\,,
\ee
where $P=P_QI_{\rm n}$ and $I_{\rm n}=I/3MR^2$; $I_{\rm n}$ is
the dimensionless modal inertia plotted in Fig.~3.
In the lower panel of Fig.~6, we plot the quantity $\lambda^2P$.
All the quantities plotted in this figure are applicable also to
nonradial modes of low degree.
However, for nonradial modes we have to take
into account the damping effects in the g-mode propagation zone.
With the help of equation~(21) and the data given in Fig.~3 we can
evaluate amplitudes for radial modes with $D_{\rm p}<0$.

In Fig.~7, we compare radial-mode frequencies and amplitudes
calculated for M$_\alpha$ with the observational data of $\alpha\,$UMa.
Keeping in mind the large observational errors and the fact that
we have made no effort to adjust model parameters to fit the
frequencies, we regard the agreement of frequencies as
satisfactory. On the other hand, the disagreement between the amplitudes
is very serious: the observed amplitude at $n=1$ exceeds the
predicted value by three orders of magnitude, and the frequency dependence
of the amplitudes differ drastically.

\begin{figure}
\epsfxsize=8cm \epsffile{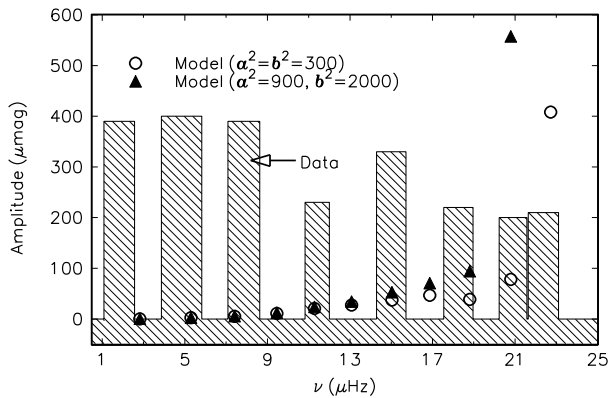} \caption{ Amplitudes and
frequencies of oscillations in $\alpha\,$UMa from Buzasi\et(2000)
compared with model calculations obtained for two sets
of convection parameters. The width of the shadowed rectangles
corresponds to the uncertainty of the frequency data. Calculated
amplitudes for the $n=1$, 2, and 3 modes are 0.5, 2.7, and 5.8
$\mu$mag, respectively. 
The observed peaks at $\nu=34.9\pm0.6$ and $\nu=43.6 \pm0.9\,\mu$Hz, 
which are above the calculated acoustic cut-off frequency, are not shown.}
\end{figure}

An additional difficulty is presented by the presence of the two peaks
above the acoustic cut-off frequency. Such high-frequency peaks are observed
in the Sun, but with amplitudes much lower than those below the
acoustic cut-off. The two highest-frequency peaks in $\alpha\,$UMa have
amplitudes of about 0.2\,mmag, which are similar to most of the other peaks.
We should stress that the amplitude estimates in Fig.~7 were obtained
using the pulsation modes of a model with an atmosphere based on model C
of Vernazza, Avrett \& Loeser (1981). That atmosphere has an acoustic 
cut-off frequency of $32.4 \mu$Hz at the temperature minimum, which is 
lower than the two highest frequencies of the observed peaks.
A more realistic atmosphere might have a higher value than this, which
itself is higher than the value for the Eddington grey atmosphere quoted 
in Table~2 and used by Guenther\et(2000).

The amplitudes of stochastically excited nonradial modes, including
those that are most efficiently trapped in the acoustic cavity, are
expected to have values much lower than those of corresponding radial modes.
Equation (20) applies to nonradial modes if the contributions to $I_{\rm n}$
and $D$ from the g-mode propagation zone are included. The values of both
$I_{\rm n}$ and $D_{\rm g}$ are substantially larger than those
plotted in the lower panels of Fig.~2 and 3, owing to the damping effect
in the outer layers.

\section{Solar-like or Mira-like oscillations ?}

The results of the stability analysis presented in Section 4.1 seem to
exclude an interpretation of the low-frequency part of the $\alpha\,$UMa
oscillation spectrum in terms of self-excited modes. Indeed, the
damping effect of convection exceeds by a large margin the driving
effect of the opacity perturbation. However, there still seems to be a
greater chance for an interpretation in terms of Mira-like excitation
than in terms of solar-like excitation. The trend of calculated
amplitudes is determined mainly by the factor $I_{\rm n}^{-1/2}$ in
equation~(21), which is the most reliably calculated quantity in the
expression. One may contemplate that for the first three modes
$D_{\rm p}$ is really much lower than what we calculated, but this option
would require near cancellation of damping and driving effects; it is more
plausible that the quantity is less than zero and that the modes are
unstable. The option that remains is an increase of $\lambda^2E$
by four to six orders of magnitude.

Whatever is the correct answer, the required changes are bound to be
related to the way in which we treat the interaction between pulsation and
convection. Our treatment, like most of those that have been used, is based
on the mixing-length formalism, and we know that it is an inadequate tool
for describing the mean properties of convection. In studies of
acoustic mode damping and excitation we have to consider more
detailed aspects of the dynamics of convection. The alternative is a
hydrodynamical simulation. This has already been applied to solar radial
oscillations by e.g. Stein and Nordlund (2001). It is to be hoped that
before long this approach will become applicable to red-giant oscillations too.

How could future observational work on red-giant variability help us?
One possibility is the disproof of genuine pulsations in $\alpha\,$UMa
and in other red giants. Short-term variability could be a direct
manifestation of convection, like large-scale granulation. This would result in
progress being slow: compare how much we have learned in the past from the
Sun's granulation with what we have learned from its oscillations.

One very promising observational approach to the solar-like
{\it vs} Mira-like alternatives is repeating the analysis of
Cook's\et(1997) MACHO data with a much longer time base or of
extensive data from another microlensing projects such as OGLE (Udalski\et1997).
Much improved frequency and amplitude resolution is expected.
Showing that the ridges extend from a few days to hundreds of days with
a continuous amplitude increase might strengthen the Mira-like interpretation.

Other observational evidence supporting a Mira-like interpretation
would be the identification of nonradial modes. We have seen in Section~3 that
if radial modes are unstable some $\ell=1$ and 2 modes should be unstable too.
If the modes are stable, then, as we discussed at the end of Section~4.2,
nonradial modes will be excited stochastically, but
their amplitudes will be much lower than those of their radial counterparts.

It could be possible that low-order modes in $\alpha\,$UMa are
Mira-like whilst those of higher order are solar-like. This could also
be the case for the multiperiodic M-type giants found by Koen and Laney (2000).
In some of these stars the frequency ratio exceeds 10, and there is
no doubt that the highest frequencies exceed the acoustic cut-off frequency.
Two of the peaks in $\alpha\,$UMa, as we have already noted, are also
above the acoustic cut-off frequency, but that does not necessarily produce
pulsational stability (cf. Balmforth\et2001)

Regardless of what the excitation mechanism is, the data on normal-mode
frequencies will be very useful as a constraint on stellar
parameters and models. Prospects of detecting nonradial modes is
particularly interesting in this context.

\section*{Acknowledgements}

Most of this work was carried out while WAD was a Raymond and Beverly Sackler
Foundation Astronomer at the Institute of Astronomy, Cambridge. Research of
WAD and RS is supported in part by the Polish grant KBN 5P03D 030 20. GH is
grateful for the support of the UK Particle Physics and Astronomy Research
Council.

\end{document}